\documentclass[12pt,onecolumn,journal]{IEEEtran}
\usepackage[encapsulated]{CJK}
\usepackage[T1]{fontenc}
\usepackage[latin9]{inputenc}
\usepackage{amsmath}
\usepackage{amsthm}
\usepackage{amssymb}
\usepackage{graphicx}

\makeatletter
\theoremstyle{plain}
\newtheorem{thm}{\protect\theoremname}
\theoremstyle{definition}
\newtheorem{defn}[thm]{\protect\definitionname}
\theoremstyle{plain}
\newtheorem{lem}[thm]{\protect\lemmaname}
\theoremstyle{plain}
\newtheorem{cor}[thm]{\protect\corollaryname}

\providecommand{\corollaryname}{Corollary}
\providecommand{\definitionname}{Definition}
\providecommand{\lemmaname}{Lemma}
\providecommand{\theoremname}{Theorem}

\begin{document}

\title{Rigorous and Generalized Proof of Security of Bitcoin Protocol with
Bounded Network Delay}
\author{Christopher Blake, Chen Feng, Xuechao Wang, Qianyu Yu}
\maketitle
\begin{abstract}
A proof of the security of the Bitcoin protocol is made rigorous,
and simplified in certain parts. A computational model in which an
adversary can delay transmission of blocks by time $\Delta$ is considered.
The protocol is generalized to allow blocks of different scores and
a proof within this more general model is presented. An approach used
in a previous paper that used random walk theory is shown through
a counterexample to be incorrect; an approach involving a punctured
block arrival process is shown to remedy this error. Thus, it is proven
that with probability one, the Bitcoin protocol will have infinitely
many honest blocks so long as the fully-delayed honest mining rate
exceeds the adversary mining rate. This means that an adversary cannot censor future transactions of a user in perpetuity, which would render the protocol useless.
\end{abstract}

\section{Introduction}

In 2009, Satoshi Nakamoto introduced the Bitcoin protocol \cite{Nakamoto_bitcoin},
a permissionless distributed ledger whose security is based on honest
nodes having over $50\%$ of hashing power. Nakamoto proved that such
a protocol was secure against the private double-spend attack, but
future papers recognized there are other possible attacks (including,
in particular, selfish mining attacks \cite{majorityIsNotEnough} and balance attacks \cite{Natoli2016TheBA}). 

We consider a generalization of the Bitcoin protocol in which multiple hashing algorithms may be used to mine the next block, like those algorithms used in existing blockchains \cite{myriadcoinHome, digibyte, verge}, and also discussed in a companion paper for an algorithm called Merged Bitcoin \cite{mergedBitcoinPaper}.

In \cite{BackboneProtocolGaray} and \emph{ et al.} and \cite{PassAsynchronous}, proofs of the security of the bounded-delay network model are given, but the bounds are not tight.
In \cite{sompolinskySecureHighRateTransactionBTC}  the authors provide a tight bound of the security of the Bitcoin protocol, but do not prove the security against any attack.

Security of the Bitcoin protocol against \emph{all} types of attacks
was finally proven in \cite{everythingARaceAndNakamoto} and \cite{tightConsistencyBounds}. However,
the proof in \cite{everythingARaceAndNakamoto} has a small but important error in its analysis. The error
involves assuming the sequences of random variables representing the
difference in adversary chain length and honest chain length is a
random walk: in reality, it is \emph{not} a random walk, which we prove in a counterexample Appendix \ref{sec:priorFlaw}.

In work that was concurrent with \cite{everythingARaceAndNakamoto}, \cite{tightConsistencyBounds} also proves the security of
Bitcoin. However, this proof is complicated and has not been generalized to the multi-hashing resources case.

Our approach is also important for the proof of another protocol called
Merged Bitcoin which is introduced in a companion paper. Hence, for
this paper, we generalize the Bitcoin model. In our case, we allow
blocks of different types, and each of these types may have different
point values. 

We define the fully-delayed score growth rate of honest nodes as the
rate of growth of honest blocks when the blocks are subject to full
network delay $\Delta$, which we call $\lambda_{h}$. The score growth
rate of all adversary blocks is defines as $\lambda_{a}$. The security
region is then proven to be $\lambda_{a}<\lambda_{h}$. 

In order for our paper to be mostly self-contained, we reproduce some of the results of \cite{everythingARaceAndNakamoto}. However, we do this for our generalized Bitcoin model in which different types of blocks may have different point values. The paper changes the approach of \cite{everythingARaceAndNakamoto} in three significant ways. First, the proof herein is generalized to the multi-block type model, allowing the proof to be used for Merged Bitcoin \cite{mergedBitcoinPaper}. Second, instead of considering the probability that a block arriving at a time $\tau^h_j$ is a Nakamoto block (a block that stays in the chain forever), we consider the probability that an interval is a \emph{Nakamoto interval}. This avoids the complication that, conditioned on a particular time, block arrivals occuring before an honest block arrival $j$ may not be independent of the honest arrival time. Finally, we resolve the issue with the former paper involving random walks, in which an erroneous assumption about block arrivals in the prior paper is remedied with a punctured arrival process technique.

\section{Models and Definitions}\label{sec:Models-and-Definitions}

\subsection{The protocol}

We consider the Bitcoin protocol as defined in \cite{Nakamoto_bitcoin},
with one small modification. We allow miners to mine blocks of different
types, where each type of block may have a different score. Miners,
both honest and dishonest, can mine blocks of any of these types.
The score of a subchain is the total score of all the blocks that
form the chain. The fork-choice rule is to mine on the chain with
the highest score, with ties broken in an arbitrary way. This generalization
does not considerably change the analysis. However, in a companion
paper, we use this general proof for a generalization of the Bitcoin
algorithm called Merged Bitcoin. 

\subsection{The Bounded Delay Network Model}

We consider the $\Delta$-bounded delay network model\footnote{This has also been called the asynchronous network with $\Delta$-bounded
delays \cite{PassAsynchronous}, the $\Delta$-synchronous model \cite{everythingARaceAndNakamoto}.
We use the term $\Delta$-bounded delay model to emphasize that the
model is \emph{not }synchronous, but rather has a delay that could
potentially be very large. }

In this model, the adversary can delay the transmission of honest
blocks up to time $\Delta$. The adversary may also delay transmission
of its own blocks to any honest miner by any time. However, once transmitted
to a single honest miner, the dishonest miner can only delay this
information by time $\Delta$ to all the other miners. This is meant
to model a case where honest miners are constantly transmitting their
view of the blockchain to the public network.

\subsection{The Arrival Processes}

For each block type, each honest miner has a fixed block-rate for
each block type, where the arrivals of each block type is a Poisson
process. We presume that each honest miner has a small total fraction
of the mining power. Hence, in any interval of length $\Delta$, we
can presume that any honest miner does not mine more than one block.
. Let each block type have some score $c_{i}>0$. Let the combined
honest block-rate (measured in blocks per second) for all the honest
miners for blocks of type $i$ be$h_{i}$. Hence, total score growth
rate of such a process when there are no delays and no adversary blocks
is $\sum c_{i}h_{i}$.

The adversary is not subject to network delays, and we presume the
adversary can see all blocks mined by honest miners as soon as they
arrive. Let the block-rate (measured in blocks per second) for the
adversary for blocks of type $i$ be $b_{i}$. Hence, the score growth
rate of the adversary mining by itself is given by $\sum c_{i}b_{i}$.

\section{Nakamoto Blocks Stay in Chain Forever }\label{sec:Proof-of-Full-Security}

In this section we will introduce the idea of a Nakamoto interval,
which is adapted from \cite{everythingARaceAndNakamoto}. we will
prove that Nakamoto intervals have an honest block (called a Nakamoto
block) which stays in the canonical chain forever. In a following
section, we will show that there these intervals occur at any point
with probability greater than $0$.

\subsection{Fully-Delayed Growth Rate is Minimal Growth Rate }\label{sec:Security-of-LinearWithNetworkDelay}

Let us first define some notation. Borrowing from \cite{everythingARaceAndNakamoto},
we let $\mathcal{T}_{h}(t)$ be the fictitious honest tree composed
of all the honest blocks that were mined since the genesis block in
the order they arrived, subject to each of them facing a network delay
of $\Delta$. As in \cite{everythingARaceAndNakamoto}, $\mathcal{T}_{h}(0)$
is the genesis block. 

Consider the mother tree $\mathcal{T}(t)$, which is the tree of \emph{all}
blocks, honest or dishonest, public or private, that exist connected
to the genesis block. Each block of that tree has a score, which is
the score of the chain that leads to it, starting at the genesis block.
\begin{defn}
The \emph{chain score }of a block $j$ is the score of the chain that
starts from the genesis block and ends at block $j$. Using this terminology,
the fork choice rule is to mine on top of the visible block with the
highest chain score.
\end{defn}
\begin{defn}
A \emph{delay schedule} for a set of arrival times is a schedule of
delays for each honest block and adversary block. For a given honest
block mined by a miner $j$, the delay schedule is a set of delays
for each other honest miner, which is some time in $[0,\Delta]$.
For each adversary blocks, the delay schedule includes a time in $[0,\infty]$
from when the block is mined by the adversary to when it is first
broadcast to an honest block. It also includes a time in $[0,\Delta]$
for each other honest miner, that indicates how much it is delayed
to all the other honest blocks.
\end{defn}
We shall see in this section that the only delay schedule we need
to consider is the fully-delayed schedule, in which all honest blocks
are delayed by $\Delta$ to the other miners.
\begin{defn}
The \emph{fully-delayed honest chain }is the hypothetical honest chain
that is produced when no adversary blocks are added and each honest
block is delayed the maximum time $\Delta$ to all other honest blocks.
\end{defn}
\begin{lem}
\label{lem:minScoreTreeMaximalDelay}The score of the highest score
honest block grows at least as fast as the fully-delayed honest chain.
\end{lem}
\begin{IEEEproof}
The analogous lemma was proven in \cite{sompolinskySecureHighRateTransactionBTC}
for the Bitcoin case and we generalize the proof for our case. First,
let us assume that the only arrivals included are honest block arrivals
(we shall consider the case of adversary arrivals being added to the
tree later in the proof). Suppose there exists a block that was not
fully delayed to all other blocks, which we shall call a \emph{non-delayed
block}. 

Let us call the tree that contains at least one node that was not
fully delayed $\mathcal{T}_{prior}$. We shall produce a new tree,
$\mathcal{T}_{delayed}$, formed using the same blocks and same arrival
times, in which this non-delayed block is fully delayed. We shall
show that the chain score of each block in this new tree is less than
or equal to its score in $\mathcal{T}_{prior}$, for all times $t$.

Let us consider a block $j$ arriving at $\tau_{j}^{h}$ that was
not fully-delayed to future blocks. Consider all the honest blocks
that arrived within time $\Delta$ from $\tau_{j}^{h}$. Some of these
may be mined on top of block $j$. Consider a particular such block,
block $m$. If block $m$ was mined on block $j$, this means that
the chain-score of block $j$ was the highest that that miner saw
when the block was mined. Now let's consider the tree produced with
the exact same delay schedule, except block $j$ was delayed fully.
With this new delay schedule, for all blocks within $\Delta$ of block
$j$, their chain score is now either the same or less. Either these
blocks were mined on block $j$, which they now no longer can be,
or they were mined on a prior block. If they previously had been mined
on a prior block, then nothing changes by adding this delay. If they
were mined on block $j$, they now must be mined on a block (that
they can still see) with chain score less than or equal to $j$ (because
$j$ was the highest score chain that it saw previously).

Thus, by adding this delay, all blocks up to $\tau_{j}^{h}+\Delta$
have chain score less than or equal to their original score.

Now we show by induction that from this time onward, the chain score
of all blocks after this are less than or equal to their score in
$\mathcal{T}_{prior}$. First, let us consider the first block that
arrives after $\tau_{j}^{h}+\Delta$. From its viewpoint, it can only
see a subset of all blocks that arrived before it. But all blocks
that arrived before it have chain score less than or equal to their
score in $\mathcal{T}_{prior}$. This new block will be added to the
highest score visible to its miner each of which are less than or
equal to their score in $\mathcal{T}_{prior}$. Hence the score of
this block is less than or equal to its score in $\mathcal{T}_{prior}$.

Now assume that all blocks up to block $k$ have chain score less
than or equal to $\mathcal{T}_{prior}$. Then, block $k+1$ has a
view of a subset of these blocks, each of which have score less than
or equal to their score in $\mathcal{T}_{prior}$. Then, block $k+1$
will be added to one of these blocks, and thus it will have score
less than or equal to what it had in $\mathcal{T}_{prior}$.

This procedure of adding full delays to blocks that were not fully
delayed can be applied until the delay schedule has only full delays,
and this can only maintain or decrease the score of the canonical
chain.

A similar proof by induction can show that removing any adversary
block from the tree can only maintain or decrease the score of all
honest blocks. Hence, the minimum score tree is produced when honest
blocks are fully-delayed and no adversary blocks are injected into
the chain. 
\end{IEEEproof}
Using the same proof by induction method, we can also show the following
lemma:
\begin{lem}
\label{lem:RemovingHonestDecreases}Removing any honest block from
the tree can only decrease or maintain the score of the canonical
chain in the view of any honest node. 
\end{lem}

\subsection{First key result: Score growth rates are additive}
\begin{defn}
We define $S_{h}(a,b)$ to be the growth in score of a fully-delayed
honest tree starting at time $a$ and ending at time $b$.
\end{defn}
We also define:
\begin{defn}
We define $S_{min}(t)$ as the minimum score of the highest score
honest block at time $t$ that is visible to all honest miners, minimized
over all possible attack strategies.
\end{defn}
\begin{lem}
\label{lem:mineGrowsAtLeastAtDeltaShiftedSh}Consider any two times
$t_{1}$ and $t_{2}$, $t_{2}\ge t_{1}+2\Delta$. For all such times:
\begin{equation}
S_{min}(t_{2})\ge S_{h}(t_{1}+\Delta,t_{2}-\Delta)+S_{h}(t_{1}).\label{eq:SminGrowthRatebound}
\end{equation}
\end{lem}
In words, the above lemma says that the score at time $t_{2}$ is
at least the score growth of the fictional fully delayed honest chain
from $t_{1}+\Delta$ to $t_{2}-\Delta$, plus the score at $t_{1}$. 
\begin{IEEEproof}
By Lemma \ref{lem:minScoreTreeMaximalDelay}, we know that maximal
delays minimize the score of the canonical chain in the view of all
honest nodes at all times. Since network delay is at most $\Delta$,
at time $t_{2}$, all nodes will see all blocks mined up to time $t_{2}-\Delta$.
From the time $t_{1}+\Delta$ to $t_{2}-\Delta$, if no blocks arrive
in the time $[t_{1},t_{1}+\Delta]$, the score of the honest chain
will increase by $S_{h}(t_{1}+\Delta,t_{2}-\Delta)$. By Lemma \ref{lem:RemovingHonestDecreases},
we know that if blocks are deleted from a chain it can only maintain
or \emph{decrease }the score of the chain in the view of any node
in the future. Hence, the score grows \emph{at least }by $S_{h}(t_{1}+\Delta,t_{2}-\Delta)$
(the amount it would have grown if there were zero arrivals in the
$\Delta$ seconds after $t_{1}$).
\end{IEEEproof}

\subsection{Nakamoto Block Definition}

We define a few events that are properties of a set of arrival times,
and an honest block $j$.

Adapting a definition from \cite{lingRenAnalysisnakamotoConsensus},
we define: 
\begin{defn}
An honest block is a \emph{loner }if no honest blocks occur in the
time $\Delta$ before and after it.
\end{defn}
Note that all loners appear in the fully-delayed honest chain that
has no dishonest blocks added.

\begin{figure}

\includegraphics[scale=0.65]{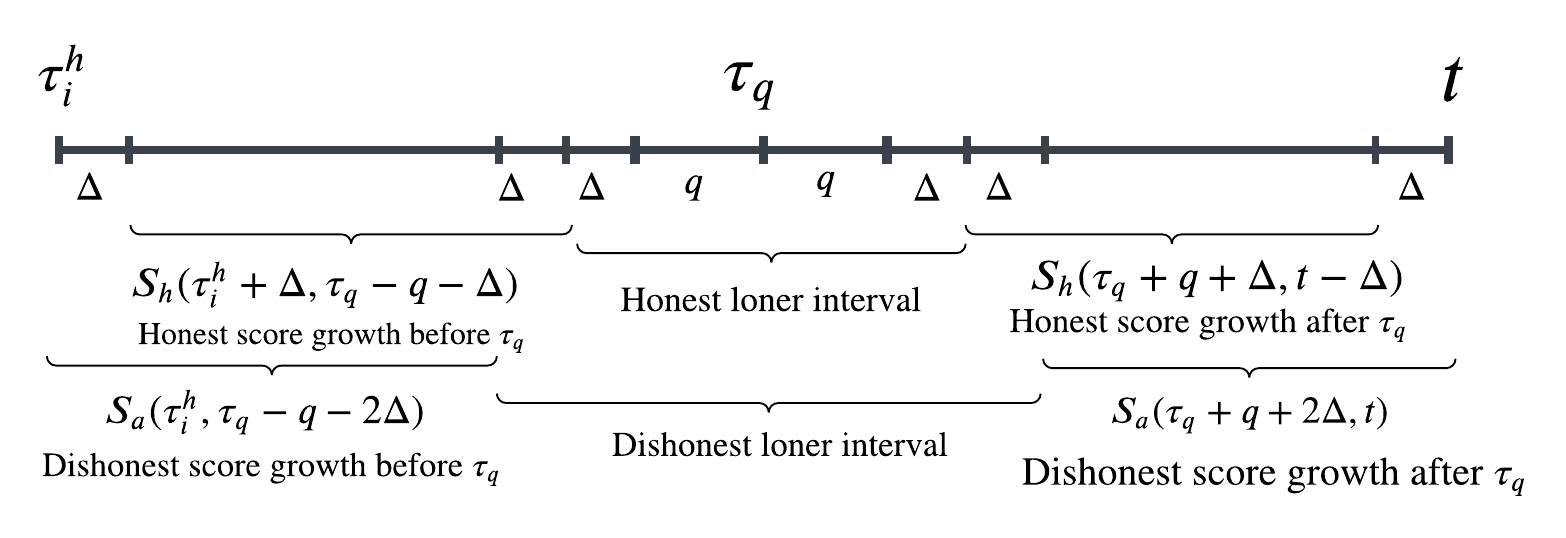}\caption{Diagram representing the past and future honest and adversary growth
intervals used in the definitions of $E_{1}$ and $E_{2}$. Observe
two things. First, the dishonest and honest growth intervals are the
same length, even though they start and end at times offset by $\Delta$.
Second, the honest score growth intervals do not overlap the honest
loner interval; similarly, the dishonest score growth intervals do
not intersect the dishonest loner interval. Hence, the event of being
a loner interval and the events $E_{1}$ and $E_{2}$ are independent.
Our proof depends on showing that with probability greater than $0$,
when in the security region, honest score growth exceeds the dishonest
score growth for all score growth intervals around a particular time
$\tau_{q}$.}\label{fig:thepastAndFutureScoreGrowthIntervals}

\end{figure}

We now adapt the notion of a Nakamoto block from \cite{everythingARaceAndNakamoto}
by considering a Nakamoto interval.
\begin{defn}
An interval of length $2q>0$ centered at a time $\tau_{q}$ is \emph{$q$-loner}
if (1) a single honest block arrives in this interval, (2) there are
no other honest blocks mined in the\emph{ honest loner interval} $[\tau_{q}-q-\Delta,\tau_{q}+q+\Delta]$,
and (3) there are no dishonest blocks in the \emph{dishonest loner
interval} $[\tau_{q}-q-2\Delta,\tau_{q}+q+2\Delta]$. The honest block
that occurs in a $q$-loner interval is called a \emph{Nakamoto block}.

We shall see why we have included the $2\Delta$ terms in the dishonest
loner interval later in the proof. Note that the block that arrives
in the length $2q$ Nakamoto interval is a loner. Hence it appears
in the hypothetical fully-delayed honest chain.

Let $L_{q}$ represent the event that a particular length $2q$ interval
is a loner interval. 
\end{defn}
We now define the arrival time properties central to being a Nakamoto
block:
\begin{defn}
\textbf{(Honest chain, ending at $\tau_{q}$, dominates from all honest
blocks in the past) :} Let $\tau_{i}^{h}$ be the arrival time of
the $i$th honest block. Let $E_{1}$ be the event that fully-delayed
honest score growth in the interval $[\tau_{i}^{h}+\Delta,\tau_{q}-\Delta-q]$
is greater than the adversary score growth in the interval $[\tau_{i}^{h},\tau_{q}-q-2\Delta]$
for all $i$ in which $\tau_{q}-q-2\Delta>\tau_{i}^{h}$:
\begin{align*}
E_{1} & :=\left[S_{h}(\tau_{i}^{h}+\Delta,\tau_{q}-q-\Delta)>S_{a}(\tau_{i}^{h},\tau_{q}-q-2\Delta)\right.\\
 & \left.\text{ for all }i\text{ such that }\tau_{q}-q-2\Delta>\tau_{i}^{h}.\right]
\end{align*}
\end{defn}
See Figure \ref{fig:thepastAndFutureScoreGrowthIntervals} for a diagram
of these intervals, which is labeled with the relevant expressions
for score growth in these intervals.
\begin{defn}
\textbf{(Honest chain, starting at $\tau_{q}$, dominates at all times
in the future)} Let $E_{2}$ be the event that the honest score growth
in the interval $[\tau_{q}+q+\Delta,t-\Delta]$ is greater than the
adversary score growth in the interval $[\tau_{q}+q+2\Delta,t]$,
for all $t>\tau_{q}+2\Delta+q$. Symbolically:
\begin{align*}
E_{2} & :=\left[S_{h}(\tau_{q}+q+\Delta,t-\Delta)>S_{a}(\tau_{q}+q+2\Delta,t)\right.\\
 & \left.\text{ for all }t>\tau_{q}+q+2\Delta\right].
\end{align*}
\end{defn}
As for the definition above, refer to Figure \ref{fig:thepastAndFutureScoreGrowthIntervals}
to see these intervals labeled, as well as the expressions for score
growth during these intervals.

We can now define a Nakamoto interval and its associated Nakamoto
block:
\begin{defn}
An interval of length $2q$ in which the event\textbf{ 
\[
L_{q}\bigcap E_{1}\bigcap E_{2}
\]
}occurs is called a \emph{Nakamoto interval}. The honest block that
arrives in that interval is called a \emph{Nakamoto block}.
\end{defn}

\subsection{Nakamoto Blocks Stay in Chain Forever Proof}

We now prove a key result of the proof of the security of the Bitcoin
algorithm: that Nakamoto blocks stay in the chain forever. This was
also proven in \cite{everythingARaceAndNakamoto}, but we reproduce
a complete proof here so that our result can be self-contained, and
also because our definition of a Nakamoto block is slightly different. 
\begin{lem}
\label{lem:ifNakamotoThenForever}If the event 
\[
L_{j}\bigcap E_{1}\bigcap E_{2}
\]
occurs for some time interval centered at $\tau_{q}$, then the block
that arrives in that interval will stay in the canonical chain forever.
\end{lem}
The key ideas are that $E_{1}$ means that at the time of the Nakamoto
block arrival, no adversary chain can dominate. $E_{2}$ implies that
at no time after the honest block arrival can the adversary dominate.
$L_{j}$ implies that the canonical chain must contain this block
because no other honest or dishonest block arrived within a time close
to it. The proof makes these intuitive observations formal.
\begin{IEEEproof}
Let $\tau_{h}$ be the time of arrival of the Nakamoto block within
the length $2q$ interval, and call this block $j$. Suppose this
block does not stay in the chain forever. Then there is the earliest
time $t^{*}\ge\tau_{h}$ in which there is a chain which dominates
the chain containing block $j$ in the view of at least one honest
node. Also, this dishonest chain has the most recent honest parent
\emph{before }honest block $j$ (which could be the genesis block). 

If the most recent honest parent of this alternate chain was \emph{not}
before honest block $j$, then either (1) it was mined on a block
that descended from $j$, or (2) it was mined on an honest block that
was produced \emph{after }block $j$ that did not descend from $j$.
In the case of (1), then this chain cannot remove block $j$ from
the canonical chain, because block $j$ is part of the chain. In the
case of (2), let $k$ be the honest block (occurring after block $j$),
upon which this dishonest chain was mined. Since block $j$ is a loner,
the miner producing block $k$ must have seen block $j$, and this
implies that there was another earlier subchain which dominated block
$j$ that was produced after block $j$ (otherwise an honest node
would not have mined on it). But we are considering the \emph{earliest
}time $t^{*}$ in which an alternate chain dominates block $j$. This
cannot occur, because then the ancestor of honest block $k$ would
be the tip of a chain that dominates the chain containing block $j$,
and $t^{*}$ would not be the earliest time that an alternate chain
dominates the chain containing block $j$.

Note that this $t^{*}$ must be after $\tau_{q}+q+2\Delta$ since
we assumed $L_{j}$, which means no adversary block arrived in the
interval $[\tau_{q}-q-2\Delta,\tau_{q}+q+2\Delta]$, and given $E_{1}$,
no adversary chain can dominate the honest chain at exactly time $\tau_{q}-q-2\Delta$.

Suppose that the adversary chain is one starting at the $i$th honest
block. By Lemma \ref{lem:mineGrowsAtLeastAtDeltaShiftedSh}, and the
fact that $j$ is a loner, the score of the tree containing block
$j$ is at least: 
\[
S_{\min}(t^{*})\ge S(\tau_{i}^{h})+S_{h}(\tau_{i}^{h}+\Delta,t^{*}-\Delta)
\]
Also, the dishonest blocks can only be mined on top of honest block
$i$ \emph{after }$i$ is produced. Hence, the increase in score of
the dishonest chain on top of block $i$ must be at most $S_{a}(\tau_{i}^{h},t^{*})$.
Recall that we are considering the earliest honest block $i$ upon
which the dominating chain was mined, and so this competing chain
must only contain adversary blocks. Thus, its score must be at most
\[
S(\tau_{i}^{h})+S_{a}(\tau_{i}^{h},t^{*}).
\]
 This means that at this time $t^{*}$, if the adversarial chain is
to replace the chain containing block $j$, it must be that: 
\begin{align}
S(\tau_{i}^{h})+S_{h}(\tau_{i}^{h}+\Delta,t^{*}-\Delta) & \le S(\tau_{i}^{h})+S_{a}(\tau_{i}^{h},t^{*})\nonumber \\
S_{h}(\tau_{i}^{h}+\Delta,t^{*}-\Delta) & \le S_{a}(\tau_{i}^{h},t^{*}).\label{eq:equationToContradict}
\end{align}
However, we have assumed $E_{1}$ and $E_{2}$ have occurred. This
means that for this particular $i$, from $E_{1}:$ 
\[
S_{h}(\tau_{i}^{h}+\Delta,\tau_{q}-q-\Delta)>S_{a}(\tau_{i}^{h},\tau_{q}-q-2\Delta)
\]
and because no adversary blocks arrived within $2\Delta+q$ of $j$:
\begin{equation}
S_{h}(\tau_{i}^{h}+\Delta,\tau_{q}-q-\Delta)>S_{a}(\tau_{i}^{h},\tau_{q}).\label{eq:E1inequality}
\end{equation}
As well, from $E_{2}$, and that no dishonest blocks arrived within
$2\Delta+q$ of $\tau_{q}$, we have that for this particular time
$t^{*}$:
\begin{equation}
S_{h}(\tau_{q}+q+\Delta,t^{*}-\Delta)>S_{a}(\tau_{q}+q+2\Delta,t^{*})=S_{a}(\tau_{q},t^{*}).\label{eq:E2Inequality}
\end{equation}
But, the growth of the fully-delayed honest chain from $(\tau_{i}^{h}+\Delta)$
to $(t^{*}-\Delta)$ is at least the sum of its growth in the interval
$[\tau_{i}^{h}+\Delta,\tau_{q}-q-\Delta]$ and the interval $[\tau_{q}+q+\Delta,t^{*}-\Delta].$
Symbolically: 
\[
S_{h}(\tau_{i}^{h}+\Delta,t^{*}-\Delta)\ge S_{h}(\tau_{i}^{h}+\Delta,\tau_{q}-q-\Delta)+S_{h}(\tau_{q}+q+\Delta,t^{*}-\Delta).
\]
Using this, and substituting the right side of this inequality with
the right sides of inequalities (\ref{eq:E1inequality}) and (\ref{eq:E2Inequality}),
we get 
\[
S_{h}(\tau_{i}^{h}+\Delta,t^{*}-\Delta)>S_{a}(\tau_{i}^{h},\tau_{q})+S_{a}(\tau_{q},t^{*})=S_{a}(\tau_{i}^{h},t^{*}),
\]
 which contradicts (\ref{eq:equationToContradict}). Hence, $E_{1}\bigcap E_{2}\bigcap L_{j}$
implies that the interval centered at $\tau_{q}$ is a Nakamoto interval
and the block $j$ that arrives during that interval stays in the
canonical chain forever.
\end{IEEEproof}

\section{Proof that a the $n$th honest block is a Nakamoto block with probability
greater than 0}

In this section we provide the main contribution of this paper. Specifically,
we show that if the adversary and honest blockrates are within the
security region which we will define, then there is an a-priori probability
greater than $0$ that any given length $2q$ interval has a Nakamoto
block. In the first subsection, we show that a quantity called the
average fully delayed growth rate, $\lambda_{h}$, exists. We show
in Appendix \ref{sec:priorFlaw} that an approach used by \cite{everythingARaceAndNakamoto}
has a flaw, specifically that a process claimed to be a random walk
is not a true random walk. In this section, we show our contribution,
which uses random walk theory on a punctured arrival process which
is, in fact, a true random walk.

Note that \cite{everythingARaceAndNakamoto} provides an alternative
proof of this theorem in Appendix C.2, which is based on the ergodic
properties of arrival times. However, the authors neither precisely
define the variables assumed to possess these properties nor specify
what those properties are, let alone prove that the variables satisfy
them. Consequently, we do not consider this alternative proof sufficient.
In this paper, we resolve these shortcomings.

\subsection{Proof That Fully Delayed Average Growth Rate Exists}

We can now proceed to show that there exists a fully-delayed average
growth rate.
\begin{defn}
\label{def:lonerDefinition}An honest block (that is not the mother
block) forms a $\Delta$-gap if no other honest blocks occur within
time $\Delta$ after this block. 

We order the $\Delta$ gaps by the time in which they arrive.

Let $T(0)$ be the time from the mother block to end of the first
$\Delta$-gap. Let $T(n)$ ($n\ge1$) be the time from the end of
$n$th $\Delta$-gap to the end of the $\left(n+1\right)$th $\Delta-$gap.
Note that these random variables have finite expected value.

Let $S(0)$ be the score of the fully delayed honest chain from the
mother block to the end of the first $\Delta$-gap. Let $S(n)$ be
the score of the fully delayed chain from the end of the $n$th to
the end of the $\left(n+1\right)$th $\Delta$-gap. Since these are
bounded by a Poisson growth process, these random variables also have
finite expected value.

\end{defn}
\begin{lem}
\label{lem:delayedGrowthRateExists}Let $S(t)$ be the score of the
fully-delayed honest chain at time $t$. Then there exists a constant
$\lambda_{h}$ in which, almost surely, $\lim_{t\rightarrow\infty}\frac{S(t)}{t}=\lambda_{h}.$
Moreover, for any $\epsilon>0$, for sufficiently large $t$, $E(\frac{S(t)}{t})\ge\lambda_{h}-\epsilon$.
\end{lem}
\begin{IEEEproof}
Note that after the arrival of the first $\Delta$-gap, the random
variables $S(n)$ and $T(n)$ are independent and identically distributed. 

First, observe the following:
\[
\lim_{t\rightarrow\infty}\frac{S(t)}{t}=\lim_{n\rightarrow\infty}\frac{S(0)+\sum_{i=1}^{n}S(i)}{T(0)+\sum_{i=1}^{n}T(i)}=\lim_{n\rightarrow\infty}\frac{\frac{S(0)}{n}+\frac{\sum_{i=1}^{n}S(i)}{n}}{\text{\ensuremath{\frac{T(0)}{n}}}\frac{\sum_{i=1}^{n}T(i)}{n}}=\frac{\lim_{n\rightarrow\infty}\frac{S(0)}{n}+\lim_{n\rightarrow\infty}\frac{\sum_{i=1}^{n}S(i)}{n}}{\lim_{n\rightarrow\infty}\frac{T(0)}{n}+\lim_{n\rightarrow\infty}\frac{\sum_{i=1}^{n}T(i)}{n}}.
\]

By the Strong Law of Large Numbers, $\lim_{n\rightarrow\infty}\frac{\sum_{i=1}^{n}S(i)}{n}$
and $\lim_{n\rightarrow\infty}\frac{\sum_{i=1}^{n}T(i)}{n}$ approach
the expected value of $S(n)$ and $T(n)$, respectively, and $\frac{S(0)}{n}$
and $\frac{T(0)}{n}$ approach $0$, all with probability $1$. Hence,
$\lim_{t\rightarrow\infty}\frac{S(t)}{t}$ approaches $\frac{E(S(i)}{E(T(i)}$,
with probability $1$. 

For the second part of the Lemma, since $\frac{S(t)}{t}$ approaches
$\lambda_{h}$ almost surely, it also does so in probability. Hence,
for any $\epsilon_{1}>0$:
\[
\lim_{t\rightarrow\infty}P\left(\frac{S(t)}{t}\le\lambda_{h}-\epsilon_{1}\right)=0.
\]
Hence, for any $\epsilon_{1}>0$ and $\epsilon_{2}>0$, there is sufficiently
large $t$, such that $P\left(\frac{S(t)}{t}\le\lambda_{h}-\epsilon_{1}\right)<\epsilon_{2}$.
Thus, 
\[
E\left(\frac{S(t)}{t}\right)\ge(1-\epsilon_{2})\left(\lambda_{h}-\epsilon_{1}\right)+\epsilon_{2}(0)=(1-\epsilon_{2})\left(\lambda_{h}-\epsilon_{1}\right)
\]
where we use the fact that for all $t$, the minimum value of $\frac{S(t)}{t}$
is $0$, for all events in the sample space. Hence, choose $\epsilon_{1}$
and $\epsilon_{2}$ to be small enough such that $(1-\epsilon_{2})\left(\lambda_{h}-\epsilon_{1}\right)>\lambda_{h}-\epsilon$,
and thus, for any $\epsilon>0$ and large enough $t$, $E\left(\frac{S(t)}{t}\right)\ge\lambda_{h}-\epsilon.$
\end{IEEEproof}
\begin{defn}
We call the constant $\lambda_{h}$ in the proof above the \emph{fully-delayed
average growth rate}, or more simply the \emph{average growth rate}. 
\end{defn}
Note that we do not actually compute what this rate is for the generalized
multiple-score model. In the standard Bitcoin case, this growth rate
is shown to be $\lambda_{h}=\frac{h}{1+\Delta h}$, where $h$ is
the average honest block-rate. If blocks can have different scores,
and blocks of each score have different growth rates, this value may
be different. We do not concern ourselves with computing these values
in this paper, but in the companion paper we provide some bounds for
this growth rate \cite{mergedBitcoinPaper}.

\subsection{The punctured arrival process}\label{sec:uncturedArrivalProcess}

\begin{figure}

\includegraphics{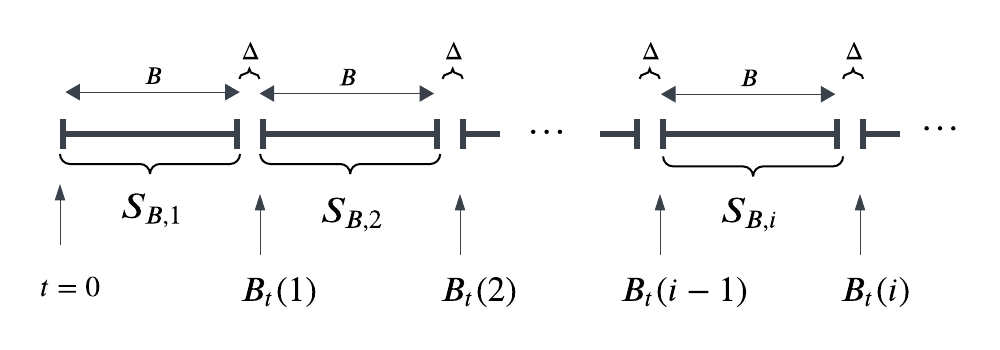}\caption{Figure representing the punctured arrival process. $B$ is the time-length
of the part of each interval that is not deleted. Without loss of
generality, we start this process at a time $t=0$. Recall that $\Delta$
is the maximum delay of the $\Delta$-delay model, and also is the
length of each puncture. In this process, all honest blocks that arrive
during these length $\Delta$ intervals are deleted. $B_t(i)$ is the\emph{
$i$th total punctured score}, which is the total score of the fully-delayed
chain after the honest blocks in the punctured intervals are deleted.
$S_{b,i}$ is the \emph{$i$th punctured interval score}, which is
the score growth of the fully-delayed honest blocks in each length
$B$ interval. Observe that, due to the length $\Delta$ deletion,
each $S_{B,i}$ is independent and identically distributed.}\label{fig:PuncturedArrivalLabel}

\end{figure}

In this section we will prove that, with probability greater than
$0$, the score growth-rate of the fully-delayed honest chain stays
close to or above average for all time, starting from any particular
time.

We shall consider a \emph{punctured} arrival process. In this process,
we run the fully-delayed delay schedule, but then, in spacings of
$B$ seconds, we \emph{puncture} the block arrivals. For simplicity,
and without loss of generality, we consider starting at time $t=0$.
In our punctured arrival process, we allow all blocks through in the
interval $[0,B]$ and then delete all blocks in the interval $[B,B+\Delta]$.
Then we let all blocks through in the interval $[B+\Delta,2B+\Delta],$
and then delete those blocks in the interval $[2B+\Delta,2B+2\Delta]$.
The process where we do this forever is called the \emph{punctured
arrival process} and the chain that this produces the \emph{punctured
chain. }See Figure \ref{fig:PuncturedArrivalLabel} for a diagram
of the punctured arrival process.
\begin{defn}
Let $S_{B}(t)$ be the score of the punctured chain at time $t$ and
call it the \emph{punctured chain score.}
\end{defn}
We also define:
\begin{defn}
Let $S(t)$ be the score of the non-punctured, fully-delayed chain,
starting at time $t=0$, called the \emph{fully-delayed score.}
\end{defn}
By Lemma \ref{lem:RemovingHonestDecreases}, deleted blocks can only
reduce the score, hence:
\begin{equation}
S(t)\ge S_{B}(t).\label{eq:initialSandSbBound}
\end{equation}
At the end of the $n$th punctured section, the time is $t=Bn+n\Delta$.
Hence, we introduce a new random process:
\begin{defn}
Let $B_t(n)=S_{B}(nB+n\Delta)$, which is a random variable that represents
the score of the punctured chain at the end of the $n$th puncture,
which we call the \emph{$n$th total punctured score.}
\end{defn}
We also define:
\begin{defn}
Let $S_{B,i}$ be the random variable representing the score of the
fully-delayed honest chain produced in the length $B$ interval $i$,
and call this the \emph{$i$th punctured interval score.}
\end{defn}
Due to the $\Delta$-gaps, and the fact that these are properties
solely of arrival times within these intervals, each $S_{B,i}$ is
independent and identically distributed. Also due to the $\Delta$-gaps,
the honest miners will see all the chain produced in the prior intervals.

Hence, the $n$th total punctured score is given by:
\[
B_t(n)=\sum_{i=1}^{n}S_{B,i}.
\]

Using the above equation, and substituting the definition of $B_t(n)$
into (\ref{eq:initialSandSbBound}) gives us, for each integer $n>0$:
\[
S(Bn+\Delta n)\ge B_t(n)=\sum_{i=1}^{n}S_{B,i}
\]
Hence, the non-punctured score at the end of each interval is at least
the sum of all the punctured interval scores up to that point. 
\begin{lem}
\label{lem:SbisStayCloseToOrAboveAverageAlways}For any $\epsilon>0$,
and for all sufficiently large $B$,
\[
P\left(\left[\frac{B_t(n)}{n}\ge B\lambda_{h}-B\epsilon\right]\text{ for all \ensuremath{n>0}}\right)
\]
is some number $p>0$. 
\end{lem}
\begin{IEEEproof}
First, due to Lemma \ref{lem:delayedGrowthRateExists},
we can choose $B$ such that $E(S_{B,i})>B\lambda_{h}-B\epsilon$
for this $\epsilon$. We let 
\[
Z_{B}[n]=\sum_{i=1}^{n}\left(S_{B,i}-\left(B\lambda_{h}-B\epsilon\right)\right)
\]
be a random walk. Note that it is a random walk because each of the
$S_{B,i}$s are independent and identically distributed. The expected
drift of this random walk is greater than $0$ since we have chosen
$B$ so that $E(S_{b,i})>B\lambda_{h}-B\epsilon$. Thus, by random
walk theory, there is a probability greater than $0$ that this random
walk goes above $0$ after the first step and then always stays above
$0$. Which means there is a probability greater than $0$ that 
\[
\sum_{i=1}^{n}\left(S_{B,i}-\left(B\lambda_{h}-B\epsilon\right)\right)>0\text{ for all integers \ensuremath{n>0}}
\]
This event is equivalent to:
\[
\sum_{i=1}^{n}S_{B,i}>\sum_{i=1}^{n}\left(B\lambda_{h}-B\epsilon\right)\text{ for all integers \ensuremath{n>0}}
\]

Substituting the definition of $B_t(n)$ on the left and summing up
$n$ constants on the right gives us: 
\[
B_t(n)>Bn\lambda_{h}-nB\epsilon\text{ for all integers \ensuremath{n>0}}
\]
has probability $p>0$. Dividing both side by $n$ proves the lemma.
\end{IEEEproof}
For the lemma below, recall that $S(t)$ is the score of the fully
delayed, unpunctured chain at time $t$
\begin{lem}
\label{lem:forTAtPoints}For any $\epsilon>0$, there is a $B$ sufficiently
large, so that for each point in time $t_{n}=Bn+\Delta n$ ($n>0$,
$n\in\mathbb{N}$),\textbf{ }there is a probability $q>0$ such that
$\frac{S(t_{n})}{t_{n}}>\lambda_{h}-\epsilon$. 
\end{lem}
\begin{IEEEproof}
First, choose a sufficiently small $\epsilon_{1}$ and a $B_{1}$
so that 
\[
\frac{\Delta\lambda_{h}}{B_{1}+\Delta}+\frac{B_{1}\epsilon_{1}}{B_{1}+\Delta}<\epsilon.
\]
 This is possible because because the left of the inequality approaches
$0$ as $\epsilon_{1}$approaches $0$ and $B_{1}$ approaches infinity.
Then, choose a $B_{2}$ such that for this $\epsilon_{1}$:
\[
E(S_{B_{2},i})>B_{2}\lambda_{h}-B_{2}\epsilon_{1}
\]
(which is possible because of Lemma \ref{lem:SbisStayCloseToOrAboveAverageAlways}):
Finally, let $B=\max(B_{1,}B_{2})$ so that:
\begin{equation}
\frac{\Delta\lambda_{h}}{B+\Delta}+\frac{B\epsilon_{1}}{B+\Delta}<\epsilon\label{eq:theBoundFromBandEpsilon}
\end{equation}
and 
\[
E(S_{B,i})>B\lambda_{h}-B\epsilon_{1}.
\]

Because of (\ref{eq:initialSandSbBound}),\textbf{ }Lemma \ref{lem:SbisStayCloseToOrAboveAverageAlways},
and $Bn=t_{n}-\Delta n$, with probability $p>0$, at each $t_{n}=Bn+\Delta n$,
for our choice $B$ and $\epsilon_{1}$\textbf{:
\begin{align*}
S(t_{n}) & \ge\sum_{i=1}^{n}S_{B,i}\ge Bn\lambda_{h}-nB\epsilon_{1}=(t_{n}-\Delta n)\lambda_{h}-nB\epsilon_{1}\\
 & =t_{n}\lambda_{h}-\Delta n\lambda_{h}-nB\epsilon_{1}.
\end{align*}
} We divide by $t_{n}=Bn+\Delta n$: 
\[
\frac{S(t_{n})}{t_{n}}>\lambda_{h}-\frac{\Delta n\lambda_{h}}{Bn+\Delta n}-\frac{nB\epsilon_{1}}{Bn+\Delta n}=\lambda_{h}-\left(\frac{\Delta\lambda_{h}}{B+\Delta}+\frac{B\epsilon_{1}}{B+\Delta}\right)>\lambda_{h}-\epsilon
\]
where in the final inequality we use our choice of $\epsilon_{1}$and
$B$ so that $\left(\frac{\Delta\lambda_{h}}{B+\Delta}+\frac{B\epsilon_{1}}{B+\Delta}\right)<\epsilon$
from (\ref{eq:theBoundFromBandEpsilon}). 
\end{IEEEproof}
\begin{lem}
There is a probability $q>0$ such that \label{lem:sometimesHonestArrivalsStayCloseToOrAboveAverage}
\[
\frac{S(t)}{t}>\lambda_{h}-\epsilon
\]
for all $t$ after the first arrival and for any $\epsilon>0$.
\end{lem}
\begin{IEEEproof}
First we show that this is true for all $t$ sufficiently large. We
know from Lemma \ref{lem:forTAtPoints}, that at points in time $t_{n}=Bn+\Delta n$,
$(n>0,n\in\mathbb{N})$, and any $\epsilon_{1}>0$, that there is
some probability greater than $0$ that: 
\[
\frac{S(t_{n})}{t_{n}}>\lambda_{h}-\epsilon_{1}.
\]
Choose $\epsilon_{1}=\frac{\epsilon}{2}$.

We assume pessimistically that this occurs but that all increases
in $S(t)$ occur exactly at the times $t_{n}=Bn+\Delta n$. Thus,
for all time in the interval $B+\Delta$ after the $t_{n}$s, the
honest chain has no arrivals. But, this allows us to assume that 
\[
S(t)\ge(Bn+\Delta n)\lambda_{h}-(Bn+\Delta n)\frac{\epsilon}{2}
\]
for all $t\in[Bn+\Delta n,B(n+1)+\Delta(n+1)]$. Dividing by $t$
on the left side, and dividing by the maximum value of $t$ in each
interval on the right side, we conclude that for all $t$ in this
interval: 
\begin{align*}
\frac{S(t)}{t}\ge\frac{(Bn+\Delta n)\lambda_{h}-(Bn+\Delta n)\frac{\epsilon}{2}}{B(n+1)+\Delta(n+1)} & =\\
\frac{(Bn+\Delta n)(\lambda_{h}-\frac{\epsilon}{2})}{(n+1)(B+\Delta)}\ge\frac{n(B+\Delta)(\lambda_{h}-\frac{\epsilon}{2})}{(n+1)(B+\Delta)} & =\frac{n}{n+1}\left[\lambda_{h}-\frac{\epsilon}{2}\right].
\end{align*}
Observe that $\frac{n}{n+1}$ approaches $1$ from below, so for sufficiently
large $n$, this expression can be made as close as possible to $\lambda_{h}-\frac{\epsilon}{2}$.
So choose $n$ so that it is within $\frac{\epsilon}{2}$ of $\lambda_{h}-\frac{\epsilon}{2}$.
Hence, with probability greater than $0$, for any $\epsilon>0$,
there is sufficiently large $n$, (and thus for all $t$ greater than
some $t'$), such that:
\begin{equation}
\frac{S(t)}{t}>\lambda_{h}-\frac{\epsilon}{2}-\frac{\epsilon}{2}=\lambda_{h}-\epsilon.\label{eq:BoujndForSufficientlyLarget}
\end{equation}

Now we prove this is true for all $t$ after the first arrival.

Let $t'$ be the minimum $t$ for which the inequality \ref{eq:BoujndForSufficientlyLarget}
holds. To show that the lemma holds for \emph{all $t$, }we simply
need to show that $\frac{S(t)}{t}>\lambda_{h}-\epsilon$ for all $t$
after first arrival in $[t,t'${]}. Note that when $\epsilon$ is
fixed, $t'$ is a fixed constant. Hence, the probability $\frac{S(t)}{t}>\lambda_{h}-\epsilon_{3}>0$
after the first arrival and until $t'$ is simply some constant greater
than $0$. Combining this with the argument above for $t>t'$, and
the lemma follows.
\end{IEEEproof}
\begin{lem}
\label{lem:adversaryStaysCloseToOrBelowAverage}With probability greater
than $0$, the number of adversary block arrivals stays close to or
below average. That is, with probability greater than $0$, for any
$\epsilon>0$ and $t_{2}>t_{1}:$
\[
S_{a}(t_{2})-S_{a}(t_{1})\le(\lambda_{a}+\epsilon)(t_{2}-t_{1}).
\]
\end{lem}
\begin{IEEEproof}
Without less of generality, let $t_{1}=0$. Divide time into small
intervals of time $\delta$. In each interval the average number of
adversary arrivals is $\lambda_{a}\delta$. Define $X[n]$ as the
adversary score in the $n$th such interval, and $S[n]=\sum X[n]$
as the score at time $t=\delta n$. Let $\epsilon>0$ and define: 

\[
Z[n]=\sum\left[X[n]-\left(\lambda_{a}+\epsilon\right)\delta\right]
\]

Note that $Z[n]$ forms a random walk since each $\left[X[n]-\left(\lambda_{a}+\epsilon\right)\delta\right]$
is independent and identically distributed. Finding the expected value
of each step of the random walk gives us:
\begin{align*}
E(\left[X[n]-\left(\lambda_{a}+\epsilon\right)\delta\right]) & =\lambda_{a}\delta-\lambda_{a}\delta-\epsilon\delta\\
 & =-\epsilon\delta<0.
\end{align*}
Hence, the random walk has negative drift, and by random walk theory,
with probability greater than $0$, $Z[n]$ becomes negative and never
returns to $0$. Hence, with probability greater than $0$, for all
$n>0$:

\begin{align*}
\sum\left[X[n]-\left(\lambda_{a}+\epsilon\right)\delta\right] & <0\\
S[n] & <\left(\lambda_{a}+\epsilon\right)\delta n
\end{align*}

But $S[n]$ is just $S_{a}(t)$ for $t=\delta n$. The lemma is then
implied by taking the limit as $\delta$ approaches $0$.
\end{IEEEproof}

\subsection{Using this to bound the event probabilities}
\begin{lem}
\label{lem:E2GreaterThan0}If $\lambda_{h}>\lambda_{a}$, then for
any honest arrival time $\tau_{q}$, there is some chance greater
than $0$ that event $E_{2}$, rewritten below, occurs:
\begin{equation}
E_{2}:=\left[S_{h}(\tau_{q}+q+\Delta,t-\Delta)>S_{a}(\tau_{q}+q+2\Delta,t)\text{ for all }t>\tau_{q}+q+2\Delta\right].\label{eq:theCondition}
\end{equation}
\end{lem}
\begin{IEEEproof}
We rewrite our condition as: 
\begin{equation}
\lambda_{h}-\frac{\epsilon}{2}>\lambda_{a}+\frac{\epsilon}{2}.\label{lambdaHGreaterThanLambdaAwithanEpsilon}
\end{equation}
for some $\epsilon>0$. We shall use this $\epsilon$ in the remainder
of the proof.
\end{IEEEproof}
Let $A_{1}$ be the event that the honest, fully-delayed chain score
stays within $\frac{\epsilon}{2}$ or above average for all time greater
than $\tau_{q}+q+\Delta$. Symbolically, this is:
\begin{align}
A_{1} & =S_{h}(\tau_{q}+q+\Delta,t-\Delta)\ge(t-\tau_{q}-q-2\Delta)(\lambda_{h}-\frac{\epsilon}{2})\nonumber \\
 & \text{ for all }t>\tau_{q}+q+\Delta.\label{eq:A1Definition}
\end{align}
Let $A_{2}$ be the event that the adversary chain has with $\frac{\epsilon}{2}$
or below average rate of arrival for all times greater than $\tau_{q}+q+2\Delta$.
\[
A_{2}=S_{a}(\tau_{q}+q+2\Delta,t)\le(t-\tau_{j}^{h}-2\Delta)(\lambda_{a}+\frac{\epsilon}{2})
\]

We note from Lemma \ref{lem:sometimesHonestArrivalsStayCloseToOrAboveAverage}
that $A_{1}$ occurs with probability greater than $0$. As well,
from Lemma \ref{lem:adversaryStaysCloseToOrBelowAverage}, $A_{2}$
occurs with probability greater than $0$.

Using \ref{lambdaHGreaterThanLambdaAwithanEpsilon} and multiplying
both sides by $(t-\tau_{q}-q-2\Delta)$:
\[
(t-\tau_{q}-q-2\Delta)(\lambda_{h}-\frac{\epsilon}{2})>(t-\tau_{q}-q-2\Delta)\left(\lambda_{a}+\frac{\epsilon}{2}\right).
\]

If $A_{1}$ occurs, then, for the left side of this inequality: 
\[
S_{h}(\tau_{q}+q+\Delta,t-\Delta)>(t-\tau_{q}-q-2\Delta)(\lambda_{h}-\frac{\epsilon}{2})
\]
.

If $A_{2}$ occurs, then, for the right side of this inequality: 
\[
(t-\tau_{q}-q-2\Delta)(\lambda_{a}+\frac{\epsilon}{2})\ge S_{a}(\tau_{q}+q+2\Delta,t).
\]

Combining these gives us
\[
S_{h}(\tau_{q}+q+\Delta,t-\Delta)>S_{a}(\tau_{q}+q+2\Delta,t).
\]

Since $A_{1}$ and $A_{2}$ each occur with probability greater than
$0$, and they are properties independent arrival processes (the adversary
and honest arrival process), their intersection also occurs with probability
greater than $0$. Since 
\[
E_{2}=A_{1}\cap A_{2},
\]
therefore $E_{2}$ also occurs with probability greater than $0$.
\begin{lem}
\label{lem:E1GreaterThan0}The event $E_{1}$ occurs with probability
greater than $0$.
\end{lem}
\begin{IEEEproof}
In this case, one can show that there is a probability greater than
$0$ that honest block arrivals stay close to or above average for
all intervals beginning before$\tau_{q}-q-\Delta$ and ending at $\tau_{q}-q-\Delta$.
This involves dividing time up in punctured intervals of size $B$,
and then recognizing the sum of the score growth in these punctured
intervals as a random walk. This can show that at the endpoints of
these intervals there is positive probability that honest score growth
rate is close to or above average. We follow the same proof as above
to show that all the time between these endpoints also have this property
with positive probability. The rest of the proof is symmetrical to
the proof above, so we omit it for conciseness. 
\end{IEEEproof}

\begin{lem}
\label{lem:Some-blocks-areNakamotoBlocks}For any $q>0$, the event
that any given interval of size $2q$ centered at a time $\tau_{q}>q+2\Delta$
is a Nakamoto interval (and thus contains an honest block that stays
in the chain forever), has probability greater than $0$. 
\end{lem}
\begin{IEEEproof}
Note $L_{q}$ occurs with probability greater than $0$, since it
is a property of arrival times in a finite interval. Event $E_{1}$
is a property of arrival times of honest blocks outside the honest
loner interval and of dishonest blocks outside the dishonest loner
interval. Hence, $L_{q}$, $E_{2}$, and $E_{2}$ are independent.
Therefore:
\[
P(\text{Nakamoto interval})=P(L_{q}\cap E_{1}\cap E_{2})=P(L_{q})P(E_{1})P(E_{2})>0.
\]
\end{IEEEproof}

\section{The Bootstrap Argument}

The previous section does not actually prove that a chain will have
honest blocks if $\lambda_{h}>\lambda_{a}$. It merely states that
the $n$th block arrival will be in the chain forever with probability
greater than $0$. However, it does not prove that these events are
independent, and thus it may be that with probability greater than
$0$ (but less than $1$) there are no honest blocks in the canonical
chain. However, in this section we follow the proof in \cite{everythingARaceAndNakamoto}
to show that the probability that there are not Nakamoto blocks in
an interval of length $t$ scales exponentially to zero in length
$t$. We simplify the approach and use an induction argument, where
the base case depends on Lemma \ref{lem:Some-blocks-areNakamotoBlocks}.
This shows that, with probability $1$, there are infinitely many
honest blocks in the canonical chain whenever $\lambda_{h}>\lambda_{a}$.

In the following section, we use the symbols $A$ and $c$ to represent
arbitrary constants greater than $0$. Hence, if used in different
expressions, they do not necessarily represent the same value. 

We first define a few terms used for this section.

\begin{defn}
The \emph{score} of a sub-chain is the total score of all the blocks
containing it, starting from the mother block, and ending at the tip.
In the Bitcoin algorithm, each block has score $1$.
\end{defn}

\begin{defn}
A subchain $D$ is said to \emph{dominate} another subchain $W$ at
a particular time $t$ if subchain $D$ has a greater score at that
time \emph{and }does not contain the tip of subchain $W$.
\end{defn}
\begin{defn}
\textbf{Conflicted:} If block is not a loner it is said to be \emph{conflicted.}
\end{defn}
\begin{defn}
\textbf{Overtakable}: A target block is \emph{overtakable} if there
exists a set of arrival times of adversary blocks and a prior honest
block, such that if the dishonest blocks formed a chain starting at
the prior honest block, it would dominate the fictional fully-delayed
honest chain containing the target block that started at the prior
honest block.
\end{defn}
\begin{defn}
\textbf{Insecure: }A block that is either conflicted or overtakable
is considered \emph{insecure.}
\end{defn}
\begin{defn}
\textbf{Secure:} A block which is an honest block that is not conflicted
and is not overtakable by a dishonest subchain, is called \emph{secure.
} Note that Nakamoto blocks are necessarily secure.
\end{defn}
Note that overtakable does not mean \emph{overtaken }by a particular
adversary strategy.
\begin{defn}
The \emph{time-length }of an adversary chain is the time since the
most recent honest block (including potentially the mother block)
on the adversary chain.
\end{defn}
\begin{defn}
Let $B_{a,b}$ be the event that there are no Nakamoto blocks that
arrive in the time interval $[a,b]$, in which $0<a\le b$.
\end{defn}
We shall now consider a hypothetical interval $[s,s+t]$, $s,t>0$.
\begin{defn}
Let $B_L$ be the event that there is at least one honest block in the
interval $[s,s+t]$ that is overtakable by an adversary chain of time-length
at least $\sqrt{t}$.
\end{defn}
\begin{lem}
The probability of event $B_L$ scales to zero at least as fast as $A\exp(-c\sqrt{t})$.
More generally, the probability that there is at least one block in
a fixed interval of length $s$ that is overtakable by a length $t$
or longer adversary chain is less than $A\exp(-ct)$.\label{lem:ChernoffBoundArgument}
\end{lem}
\begin{IEEEproof}
This flows from an application of the Hoeffding Inequality on the punctured honest score growth and a Chernoff bound on the adversary score growth. The details of the argument
are given in Appendix \ref{appendixChernoff}.
\end{IEEEproof}
Observe from elementary probability theory that:
\begin{align}
B_{s,s+t} & =\left(B_L\cap B_{s,s+t}\right)\bigcup\left(B_L^{C}\cap B_{s,s+t}\right)\label{eq:NoNakamotoEventBigAnd}\\
 & \subseteq B_L\bigcup\left(B_L^{C}\cap B_{s,s+t}\right)\nonumber 
\end{align}
where the second line arises from weakening the condition in the first
term. Note that we use the notation that $B^C$ corresponds to the complement of the event $B$.

Now consider the event $\left(B_L^{C}\cap B_{s,s+t}\right)$. By definition
this is the event that no single block in the interval $[s,s+t]$
is overtakable by a $\sqrt{t}$ or longer adversary chain, and all
are insecure. If all are insecure, \emph{and }none are overtakable
by a long chain, this is the same event as each block is either conflicted
\emph{or }overtakable by a $\sqrt{t}$ or smaller adversary chain. 
\begin{defn}
\textbf{Locally secure: }A block that is neither conflicted nor overtakable
by a $\sqrt{t}$ or shorter time-length chain is called \emph{locally
secure.} Otherwise it is called \emph{locally insecure.}
\end{defn}
\begin{defn}
Let $Q_{i}$ be the event that honest block $i$ is overtakable by
a $\sqrt{t}$ or smaller chain.
\end{defn}
\begin{defn}
Let $C_{i}$ be the event that the $i$th honest block is conflicted.
\end{defn}
\[
\left(B_L^{C}\cap B_{s,s+t}\right)=\bigcap_{j:\tau_{j}\in[s,s+t]}\left(C_{j}\cup Q_{j}\right)
\]
 This is simply the AND over the events that the $j$th block in our
interval of interest is locally insecure.

\subsection{Bootstrap argument step one: Dividing interval in segments of length
$\sqrt{t}$}\label{subsec:TheBaseCaseArgument}

\subsubsection{Part one: Decomposing the Big AND into Independent Events and Applying
Independence}

Let us divide the interval $[s,s+t]$ into $\left\lfloor \sqrt{t}\right\rfloor $
segments of length $\sqrt{t}.$

Let $T_{i}$ be the $i$th such subinterval.

Now, the above conjunction over all honest blocks can be broken up
as an AND over blocks in each length $\sqrt{t}$ sub-interval, as
well as the blocks arriving in the remaining $\sqrt{t}-\left\lfloor \sqrt{t}\right\rfloor $
time. Symbolically: 
\begin{align*}
\left(B_L^{C}\cap B_{s,s+t}\right) & =\bigcap_{i=1}^{\left\lfloor \sqrt{t}\right\rfloor }\left[\bigcap_{j:\tau_{j}\in T_{i}}\left(C_{j}\cup Q_{j}\right)\right]\\
 & \bigcap_{j:\tau_{j}\in[\left\lfloor \sqrt{t}\right\rfloor ,\sqrt{t}-\left\lfloor \sqrt{t}\right\rfloor ]}\left(C_{j}\cup Q_{j}\right)
\end{align*}
where we note that $\bigcap_{j:\tau_{j}\in T_{i}}\left(C_{j}\cup Q_{j}\right)$
is an AND over all honest arrival times $\tau_{j}$ in the interval
$T_{i}$. 

But this is a subset of the AND over every third sub-interval: 
\[
\left(B_L^{C}\cap B_{s,s+t}\right)\subseteq\bigcap_{k=1}^{\left\lfloor \frac{\left\lfloor \sqrt{t}\right\rfloor }{3}\right\rfloor }\left[\bigcap_{j:\tau_{j}\in T_{3k-1}}\left(C_{j}\cup Q_{j}\right)\right].
\]
Let $N_{k}$ be the event $\bigcap_{j:\tau_{j}\in T_{3k-1}}\left(C_{i}\cup Q_{i}\right)$
(this is the event in the square brackets in the expression above).
This is the event that the $(3k-1)$\emph{th} subinterval has all
blocks locally insecure. In other words, it is the event that all
loners in the $(3k-1$)th subintervals are overtakable by an adversary
chain of length less than $\sqrt{t}$.

Then we can rewrite the expression above as as:
\[
\left(B_L^{C}\cap B_{s,s+t}\right)\subseteq\bigcap_{k=1}^{\left\lfloor \frac{\left\lfloor \sqrt{t}\right\rfloor }{3}\right\rfloor }\left[N_{k}\right].
\]

Note that we use of the floor function in the expression $\left\lfloor \frac{\left\lfloor \sqrt{t}\right\rfloor }{3}\right\rfloor $
because $\frac{\left\lfloor \sqrt{t}\right\rfloor }{3}$ may not be
divisible by $3$. Now we can go back to Expression \ref{eq:NoNakamotoEventBigAnd}:
\[
B_{s,s+t}\subseteq B_L\bigcup\left[\bigcap_{k=1}^{\left\lfloor \frac{\left\lfloor \sqrt{t}\right\rfloor }{3}\right\rfloor }\left[N_{k}\right]\right]
\]

This implies:
\begin{equation}
P\left(B_{s,s+t}\right)\le P(B_L)+P\left(\bigcap_{k=1}^{\left\lfloor \frac{\left\lfloor \sqrt{t}\right\rfloor }{3}\right\rfloor }\left[N_{k}\right]\right).\label{eq:BsandSplusTboundedByTheAndOfTheNks}
\end{equation}

Note that each $(3k-1)$th subinterval represented in $N_{k}$ has
a subinterval of length $\sqrt{t}$ before and after it that is not
neighboring the subinterval of a different $N_{k}$. Thus, each $N_{k}$
is solely a property of honest and adversary arrival times in its
own or neighboring subintervals. Therefore, these probabilities are
independent and identically distributed. Hence: 
\begin{equation}
P\left(B_{s,s+t}\right)\le P(B_L)+P(N_{1})^{\left\lfloor \frac{\left\lfloor \sqrt{t}\right\rfloor }{3}\right\rfloor }.\label{eq:PofBandN1tothesqrtT}
\end{equation}

If all blocks in a subinterval are locally insecure, then none are
secure. Thus, the event that all blocks are locally insecure is a
subset of the event that none are secure. But we know that the probability
that there are no secure blocks in an interval of length at least
$q>0$ is less than $1$ (From Lemma \ref{lem:Some-blocks-areNakamotoBlocks}).
Hence, $P(N_{1})=\rho$ for some $\rho<1$, and hence $P(N_{1})^{\left\lfloor \frac{\left\lfloor \sqrt{t}\right\rfloor }{3}\right\rfloor }=\rho^{\left\lfloor \frac{\left\lfloor \sqrt{t}\right\rfloor }{3}\right\rfloor }\le A\exp(-c(\sqrt{t}))$
for arbitrary constants $A$ and $c$.

From Lemma \ref{lem:ChernoffBoundArgument}, $P(B_L)\le A\exp(-c\sqrt{t})$
for arbitrary constants $A$ and $c$. 

Combining these two observations with expression \ref{eq:PofBandN1tothesqrtT}
above gives us:
\begin{align*}
P\left(B_{s,s+t}\right) & \le P(B_L)+P(N_{1})^{\left\lfloor \frac{\left\lfloor \sqrt{t}\right\rfloor }{3}\right\rfloor }\\
 & \le A\exp(-ct^{1/2}).
\end{align*}

\subsection{Step 2: The Induction Argument}

We have shown above that for $k=1$:
\[
P\left(B_{s,s+t}\right)\le A\exp(-ct^{1/2}).
\]

We shall assume our induction hypothesis that for some integer $k$,
for $t$ sufficiently large, 
\begin{equation}
P(B_{s,s+t})\le A\exp(-ct^{\left(\frac{k}{k+1}\right)}).\label{eq:InductionHypothesis}
\end{equation}

Note that the case of $k=1$ is proven in the section above.

We shall use the the same arguments as Section \ref{subsec:TheBaseCaseArgument},
with four differences. First, we divide the length $t$ interval into
$\left\lfloor t^{\frac{1}{k+2}}\right\rfloor $ subintervals, each
of length $t^{\frac{k+1}{k+2}}$. Second, we define local insecurity
as the event that a block is either conflicted or is overtakable by
a length $t^{\frac{k+1}{k+2}}$ or smaller adversary subchain. Third,
we define $N_{q}$ as the event that all honest blocks in the $(3q-1$)th
sub-interval are locally insecure. Fourth, we define $B_L$ as the event
that there is at least one honest block in the interval $[s,s+t]$
that is overtakable by an adversary chain of time-length at least
$t^{\frac{k+1}{k+2}}$. From this we can conclude that:
\begin{equation}
P(B_{s,s+t})\le Ae^{-ct^{\frac{k+1}{k+2}}}+\left(P(N_{q})\right)^{\left\lfloor \frac{\left\lfloor t^{\frac{1}{k+2}}\right\rfloor }{3}\right\rfloor }.\label{eq:inInductionProofProbofBsandSplustBounded}
\end{equation}

If all blocks in a subinterval are locally insecure, then none are
secure. Thus, the event that all blocks in an interval are locally
insecure is a subset of the event that none in that sub-interval are
secure. But from our induction hypothesis, this probability scales
exponentially in the length of the interval raised to $\frac{k}{k+1}.$
The length of the interval is $t^{\frac{k+1}{k+2}}$and hence, it
should decrease exponentially in 
\[
\left(t^{\frac{k+1}{k+2}}\right)^{\frac{k}{k+1}}=t^{\frac{k}{k+2}}
\]
and thus:

\[
P(N_{q})\le A\exp(-ct^{\frac{k}{k+2}}).
\]
 Plugging into \ref{eq:inInductionProofProbofBsandSplustBounded}
we get:
\begin{align*}
P(B_{s,s+t}) & \le Ae^{-ct^{\frac{k+1}{k+2}}}+A\left(\exp(-ct^{\frac{k}{k+2}})\right)^{\left\lfloor \frac{\left\lfloor t^{\frac{1}{k+2}}\right\rfloor }{3}\right\rfloor }\\
 & \le Ae^{-ct^{\frac{k+1}{k+2}}}+A\exp(-ct^{\frac{k+1}{k+2}})\\
 & \le A\exp\left(ct^{\left(\frac{k+1}{k+2}\right)}\right)
\end{align*}

Therefore this bound is true for all integers $k>1$.

As $k$ approaches infinity, $\left(\frac{k+1}{k+2}\right)\rightarrow$1,
and thus
\[
P(B_{s,s+t})\le A\exp(-ct^{1-\epsilon})
\]
 for any $\epsilon>0$ and some constants $A$ and $c$ greater than
$0$. This argument proves the main theorem of our paper:
\begin{thm}
The probability that any sub-interval has no honest blocks that stay
in the chain forever goes to zero exponentially in time-length $t$
when $\lambda_{h}>\lambda_{a}.$
\end{thm}
\begin{cor}
The probability that there are infinitely many honest blocks in the
canonical chain is $1$ if $\lambda_{h}>\lambda_{a}$.
\end{cor}

\begin{IEEEproof}
Divide time into intervals, indexed by $i$, of length $t_{i}=i$.
Let $M_{i}$ be the event that interval $i$ has no honest blocks
that stay in the chain forever. We know from above that $P(M_{i})\le A\exp(ci^{1-\epsilon})$
for large enough $i$.

Observe that, since these probabilities are exponential, therefore
$\sum_{i=1}^{\infty}P(M_{i})<\infty$. By the Borel-Cantelli lemma  \cite{borelLemma, cantelliLemma},
the event that these insecure intervals occur infinitely often has
probability $0$. Hence, with probability one, there will be a time after which all these
intervals have a secure block. Hence, the canonical chain has infinitely
many honest blocks with probability $1$ when $\lambda_{h}>\lambda_{a}$.
\end{IEEEproof}
This implies something with practical significance. If there are infinitely many honest blocks with probability $1$, one can be sure that in the future there will some other honest block in the chain that can accept a user's transaction. This prevents the adversary from being able to censor all future transactions of a user, which indeed would make the protocol useless as a public ledger. This does not imply, however, that future transactions cannot be censored by overwhelming each honest block with an excess of transactions, leaving no room for a target transaction to make it into the block.
\section*{Insecurity Region}
\begin{thm}
If $\lambda_{a}>\lambda_{h}$, then Merged Bitcoin is insecure, and
there exists an attack in which the canonical chain, with $100\%$
probability has all dishonest blocks.
\end{thm}
\begin{IEEEproof}
The attack that will work is the private mining attack. The adversary
simply mines a chain in private and then reveal it whenever it has
score greater than the highest score honest block. The reader should
observe that by the law of large numbers, between every reveal, with
probability one, eventually the adversary chain will dominate the
honest chain.
\end{IEEEproof}

\section{Conclusion}
We considered a generalization of the Bitcoin protocol in which blocks of different types (mined by different types of hashing resources), have a different score. For this generalized model we have simplified and made more rigorous the Nakamoto block approach to proving security. The key differences in previous approaches are as follows. First, we considered the probability of a Nakamoto interval, which is an interval that contains a block which stays in the chain forever. Second, we defined a punctured arrival process which is a true random walk, resolving an issue with the prior approach in which a process assumed to be a random walk was not actually a random walk. Finally, we simplified the induction-based bootstrap argument to show exponential scaling in the probability of no honest forever block. This work can be used to prove security of multi-hash algorithm blockchains like those in \cite{myriadcoinHome, digibyte, verge}, which is also discussed in a companion paper for an algorithm called Merged Bitcoin \cite{mergedBitcoinPaper}.

\bibliographystyle{IEEEtran}
\bibliography{bibtextDoc}

\appendices

\section{Prior paper flaw}\label{sec:priorFlaw}

We will prove using a counterexample a flaw with the prior proof in
\cite{everythingARaceAndNakamoto}.

The authors of \cite{everythingARaceAndNakamoto} state in Appendix
C.1: ``$S[n]$ {[}represents the{]} difference between the increase
in $D_{h}$ and the number of adversary arrivals.'' In this case,
$D_{h}$ is the depth of an honest, fully-delayed chain. 

The claim is equivalent to $S[n]=S[n-1]+X_{n}$ for some $X_{n}$s
that are independent and identically distributed. This implies that
$S[n]-S[n-1]=X_{n}$. 

The authors continue: ``Hence, $S[n]$ simply counts the difference
between the number of honest and adversary arrivals when there are
$n$ arrivals in total. In this case, $S[n]$ jumps up by $1$ when
there is an honest arrival, and, goes down by $1$ when there is an
adversary arrival.''

The authors then apply the well-known result that a random walk with
positive drift leaves the origin once and never returns with probability
greater than $0$.

Based on the stated definition of $S[n]$, $X_{n}$ is a random variable
that tracks the next block in the hypothetical race between a\emph{
fully-delayed} honest chain and the adversary chain. It is defined
as:
\[
X_{n}=\begin{cases}
+1 & \text{nth block is honest}\\
-1 & \text{nth block dishonest}
\end{cases}
\]

We shall prove that these random variables are not independent. Therefore
the conclusions drawn in \cite{everythingARaceAndNakamoto} do not
flow from the arguments.

Indeed, a generalized random walk does not necessarily behave like
a simple random walk. For example, consider a process that starts
out deterministically leaving the origin, returning, and then continuing
on forever as a normal random walk with positive drift. This process
is similar to a simple random walk, and the average drift of this
process is clearly positive, but with probability one it leaves the
origin and \emph{always} returns.

We can fix this issue with a punctured arrival technique, but we shall
now prove that the $X_{n}$s defined above are not independent.

\subsection{Counterexample}

For this section, let $b$ be the dishonest mining rate and $h$ be
the honest mining rate (before there are any delays). 

We consider an honest chain growing when subject to full-delays by
the adversary, and consider the classic Bitcoin case of all blocks
being the same score.

Consider the probability that there is a dishonest arrival in the
time $\Delta$ after a particular non-orphaned honest block of the
fully-delayed honest chain. The arrivals of these dishonest blocks
is a Poisson process with parameter $b$. Applying the well-known
probability distribution of number of arrivals of a Poisson process
with parameter $b$ in time interval $\Delta$:
\[
P(\text{dishonest block in \ensuremath{\Delta} after honest block})=1-P(0\text{ blocks in time \ensuremath{\Delta})}=1-e^{-\Delta b}
\]

We are in the security region considered when:
\[
b<\frac{h}{1+\Delta h}.
\]
 Let $\Delta$ and $h$ be very large, (say, 100).

Now, in this regime, let $b$ be very close to but within the security
region boundary:
\[
b\approx\frac{h}{1+\Delta h}=\frac{1}{\frac{1}{h}+\Delta}\approx\frac{1}{\Delta}
\]
where the approximation comes from $h$ and $\Delta$ each being much
larger than 1. 

Then:
\[
P(\text{at least \ensuremath{1} dishonest block in \ensuremath{\Delta} after honest block})\approx1-e^{-\Delta\frac{1}{\Delta}}=1-e^{-1}\approx0.63
\]

Note that if any dishonest block arrives in the time $\Delta$ after
an honest block of the fully-delayed honest tree, then the first of
these blocks will be the next block. It doesn't matter if other honest
blocks arrive since they will be orphaned due to the network delay.
Thus, the probability that the next block is dishonest given that
the last block is honest is \emph{at least }the probability that a
dishonest block arrives in time $\Delta.$ 

Combining this with the approximation above, this implies that:
\[
P(X_{n}=-1|X_{n-1}=+1)\ge0.63.
\]

For the $X_{n}$s to be independent, we require that $P(X_{n}=-1|X_{n-1}=+1)=P(X_{n}=-1).$
However, since we are in a security region, the probability that the
$n$th block is dishonest must be less than $50$ percent. Thus $P(X_{n}=-1)<0.5$,
and thus the $X_{n}s$ are not independent. Hence, the sequence $S[n]$
is not a random walk and presumptions about the drift of a random
walk made in \cite{everythingARaceAndNakamoto} do not necessarily
hold.

\section{Probability that Honest Growth Rate Deviates from Average Decays Exponentially}\label{appendixChernoff}

As in previous sections, for this appendix, for simplicity of notation,
we will continue to use the convention that $A$ and $c$ are arbitrary
positive constants (and not necessarily the same constant in every
expression).

We shall prove that the event $B_L$ (that
there is at least one honest block in the interval $[s,s+t]$ that
is overtakable by an adversary chain of time-length at $t'$) scales
as $Ae^{-ct'}.$

We let the adversary score growth rate be $\lambda_{a}$. We consider
a region in which $\lambda_{h}>\lambda_{a}$ (the security region of
the protocol).

\subsection{Probability honest growth rate deviates from average is exponential
in time $t$ }

Consider the punctured process defined in Section \ref{sec:uncturedArrivalProcess}, in which $B$ is the length of each punctured interval, $S_{B,i}$ is the score growth of the fully-delayed honest chain in the $i$th such punctured interval, and $B_{t}(n)=\sum_{i=1}^{n}S_{B,i}$.

For any choice of $\epsilon>0$, from Lemma \ref{lem:delayedGrowthRateExists} we know that as $B$
gets large the expected value of  $S_{B,i}$ approaches $B \lambda_{h}$. So choose
$B$ sufficiently large so that 
\[
E(S_{B,i})=B(\lambda_{h}-\delta\epsilon)
\]
and
\begin{equation}
-B(1-\delta)\epsilon+\Delta(\lambda_{h}-\epsilon)<0.\label{eq:BSufficientlyLargeConditionBecomesNegative}
\end{equation}
The latter inequality can be true so long as $\delta$ is sufficiently
small (which occurs as $B$ gets larger) and $B$ sufficiently big.

We have the time at the end of the $n$th puncture is:
\[
t=(B+\Delta)n.
\]
We let 
\[
n(t)=\left\lfloor \frac{t}{B+\Delta}\right\rfloor 
\]
be the number of punctured intervals that have passed by time $t$. 

We additionally let $S_{p}(t)$ be the score of the punctured chain
at time $t$. 

By Lemma \ref{lem:RemovingHonestDecreases}, deletion can only decrease the score.
Also, truncating to the nearest multiple of $t$ can also only decrease
the score. Hence:
\[
B_{t}(n(t))=B_{t}\left(\left\lfloor \frac{t}{B+\Delta}\right\rfloor \right)\le S_{p}(t)\le S(t).
\]

Consider the event $S(t)\le t(\lambda_{h}-\epsilon)$, (\emph{i.e.} that the fully-delayed honest chain at time $t$ has average score
growth rate a value $\epsilon>0$ below average. 

If we weaken the honest score growth rate by puncturing it and also
truncating it at the endpoints of each of the punctured segments,
this can only \emph{increase }the probability that the score growth
is \emph{below }a certain value. Hence, 
\[
P\left[S(t)\le t(\lambda_{h}-\epsilon)\right]\le P\left[B_{t}(n)\le t(\lambda_{h}-\epsilon)\right]
\]

Note that $\left(n+1\right)(B+\Delta)\ge t$ (where implicitly $n$
is a function of $t$). Observe that replacing $t(\lambda_{h}-\epsilon)$
with a bigger value can only \emph{increase }the probability
that $B(n)$ is less than it. Hence, replacing $t$ with $\left(n+1\right)(B+\Delta)$
gives us,
\begin{equation}
P\left[S(t)\le t(\lambda_{h}-\epsilon)\right]\le P\left[B_{t}(n)\le t(\lambda_{h}-\epsilon)\right]\le P\left[B_{t}(n)\le\left(n+1\right)(B+\Delta)(\lambda_{h}-\epsilon)\right]\label{eq:chainOfInequalities}
\end{equation}

For now we consider understanding the event $\text{\ensuremath{\left\{  B_{t}(n)\le(n+1)(B+\Delta)(\lambda_{h}-\epsilon)\right\} } }$.

Then, the event above can be expressed as:
\[
\left\{ B_{t}(n)\le\left(n+1\right)(B+\Delta)(\lambda_{h}-\delta\epsilon-(1-\delta)\epsilon)\right\} 
\]
Expanding the right side and simplifying:
\[
B_{t}(n)\le nB(\lambda_{h}-\delta)\epsilon-nB(1-\delta)\epsilon+(n+1)\Delta(\lambda_{h}-\epsilon)
\]
Moving some terms to the other side of the inequality:
\[
B_{t}(n)-nB(\lambda_{h}-\delta\epsilon)\le-nB(1-\delta)\epsilon+(n+1)\Delta(\lambda_{h}-\epsilon)
\]
Dividing both sides by $n$:
\[
\frac{B_{t}(n)-n(B)(\lambda_{h}-\delta)\epsilon}{n}\le-B(1-\delta\epsilon)+\frac{(n+1)\Delta(\lambda_{h}-\epsilon)}{n}.
\]

Observe that for the right side of the inequality, 
\[
-B(1-\delta)\epsilon+\frac{(n+1)\Delta(\lambda_{h}-\epsilon)}{n}\rightarrow-B(1-\delta)\epsilon+\Delta(\lambda_{h}-\epsilon)
\]
as $n$ gets large. Thus, from \ref{eq:BSufficientlyLargeConditionBecomesNegative},
the right side is negative for sufficiently large $n$ . Let $-k$
equal the right side of this inequality. Moreover, recognize that $B_{t}(n)=\sum_{i=1}^{n}S_{B,i}$ and
$E(S_{B,i})=(B)(\lambda_{h}-\delta\epsilon)$. Thus, we are this interested
in the event that:
\[
P\left[\frac{\sum_{i=1}^{n}S_{B,i}-\mu_{p}}{n}\le-k\right].
\]

The Hoeffding inequality states that for random variables $X(n)$
that are independent and bounded in the range $[a,b]$ for some finite
$a$ and $b$, and all $t>0$:

\[
P\left[\frac{\sum_{i=1}^{n}X(i)-E(X(i))}{n}\le-t\right]\le\exp\left(-\frac{2nt^{2}}{(b-a)^{2}}\right).
\]

Observe that the value of $S_{B,i}$ is at least $0$ almost surely (corresponding
to the event of no honest arrivals at all in the interval), and at
most is $b:=\max_{i}(c_{i})\frac{B}{\Delta}$, corresponding to the
case where the highest score block type (with score $\max_{i}(c_{i})$) arrives exactly at $\Delta$
after each prior block arrival. Hence, the Hoeffding inequality applies
for $n$ sufficiently large, and thus
\[
P\left[\frac{B_t(n)-\mu_{p}}{n}\le-k\right]\le\exp\left(-\frac{2nk^{2}}{(b)}\right).
\]
Substituting this result back into the chain of inequalities in \ref{eq:chainOfInequalities},
and substituting $n=\left\lfloor \frac{t}{B+\Delta}\right\rfloor $
gives us:
\[
P\left[S(t)\le t(\lambda_{h}-\epsilon)\right]\le A\exp(-ct)
\]
for some constants $A, c>0$.

\subsection{This implies probability of a dominating event of length $t$ is
exponential}

A basic Chernoff bound argument can show that the probability that
an adversary has average growth rate above average $\lambda_{a}$
decays exponentialky to $0$ in length $t'$. Since this is just a
straightforward application of a Chernoff bound, we omit the proof
here. 

Recall we are in a region in which $\lambda_{h}>\lambda_{a}$. Hence,
for any $t'>0$:
\begin{equation}
\lambda_{h}-\epsilon>\lambda_{a}+\epsilon\label{eq:theDeltaExpressionforlambdaAandLambdah}
\end{equation}
 for sufficiently small $\epsilon$. It is now easy to see that in
order for an adversary to dominate at a time $t'$ from an honest
block arrival, it must be that either the honest blocks must grow
at a rate below $\lambda_{h}-\epsilon$ or the adversary must grow
at a rate above $\lambda_{a}+\epsilon$. Each of these occur with
exponentially decaying probability, and thus by union bound this event
also decays with exponential probability. 

\subsection{The Integration Argument: The Probability any Block in Interval is
Dominated by Long Adversary Chain}

Let $C_{t'}^{t_{s}}$ be the event that there is an adversary chain
with these three properties: (1) it starts at an honest block at time
$t_{s}$ (start time), (2) it is length greater than $t'$, (time
length), and (3), it can overtake an alternate honest chain that starts
at time $t_{s}$. 

Let $H_{t_{s}}$ be the event that there is an honest block that arrives
in a small interval around time $t_{s}$.

Once we have the exponential scaling for one event, bounding that
the probability of a length $t'$ or greater catch-up occurs to a
block in the interval $[s,s+t]$ is easy. Note from above that if
$\lambda_{h}>\lambda_{a}$, then:
\[
P(C_{t'}^{t_{s}}|H_{t_{s}})\le A\exp(-ct').
\]

As well, $P(H_{t_{s}})\le adt$ for some constant $a$.

To find the probability that a block in the target interval is dominated
by a long adversary chain, we integrate over a number of variables,
and use the fact that the integral of an exponential function is also
an exponential function.

First, $t_{3}$ will index the possible arrival times of honest blocks
in the interval $[s,s+t${]}.

Second, let $t_{2}$ index the length of the dominating chain, which
can vary from $t'$ to $\infty$. 

Finally, let $t_{1}$ index the possible start time of the dominating
chain, which can vary from $t_{3}-t_{2}$ to $t_{3}$.

\begin{align*}
P(\text{catch up of length \ensuremath{t'} or greater of block in interval}[s,s+t]) & \le\\
\int_{s}^{s+t}\int_{t'}^{\infty}\int_{t_{3}-t_{2}}^{t_{3}}P(C_{t_{2}}^{t_{1}}|H_{t_{1}})\ensuremath{P(H_{t_{1}})}dt_{1}dt_{2}dt_{3} & \le\\
\int_{s}^{s+t}\int_{t'}^{\infty}\int_{t_{3}-t_{2}}^{t_{3}}Ae^{-ct'}adt_{1}dt_{2}dt_{3}\le Ae^{-ct'}.
\end{align*}

\end{document}